\documentclass[12pt]{article}

\usepackage[utf8]{inputenc}
\usepackage{textcomp}
\usepackage{amsmath}
\usepackage[shortlabels]{enumitem}
\usepackage{chetnew}
\usepackage{tikz}
\usepackage{amssymb,amsfonts,mathrsfs,dsfont,yfonts,bbm}\usetikzlibrary{calc}
\usepackage{xcolor}
\usepackage{braket}
\usepackage[normalem]{ulem}
\usepackage{import}
\usepackage{booktabs}
\usepackage{varioref}
\usepackage{bm}

\newcommand{\be}{\begin{equation}}
\newcommand{\ee}{\end{equation}}
\newcommand{\bea}{\begin{eqnarray}}
\newcommand{\eea}{\end{eqnarray}}

\newcommand{\CA}{\mathcal{A}}
\newcommand{\CR}{\mathcal{R}}
\newcommand{\CH}{\mathcal{H}}

\newcommand{\CD}{\mathcal{D}}
\newcommand{\CE}{\mathcal{E}}

\newcommand{\CJ}{\mathcal{J}}
\newcommand{\CB}{\mathcal{B}}
\newcommand{\CC}{\mathcal{C}}
\newcommand{\CF}{\mathcal{F}}

\newcommand{\CO}{\mathcal{O}}

\newcommand{\CI}{\mathcal{I}}
\newcommand{\CN}{\mathcal{N}}

\newcommand{\CM}{\mathcal{M}}

\newcommand*{\boxcoloro}{orange}
\makeatletter
\newcommand{\boxedo}[1]{\textcolor{\boxcoloro}{%
\tikz[baseline={([yshift=-1ex]current bounding box.center)}] \node [rectangle, minimum width=1ex,rounded corners,draw] {\normalcolor\m@th$\displaystyle#1$};}}
 \makeatother
\newcommand*{\boxcolorr}{red}
\makeatletter
\newcommand{\boxedr}[1]{\textcolor{\boxcolorr}{%
\tikz[baseline={([yshift=-1ex]current bounding box.center)}] \node [rectangle, minimum width=1ex,rounded corners,draw] {\normalcolor\m@th$\displaystyle#1$};}}
 \makeatother
\newcommand*{\boxcolorb}{blue}
\makeatletter
\newcommand{\boxedb}[1]{\textcolor{\boxcolorb}{%
\tikz[baseline={([yshift=-1ex]current bounding box.center)}] \node [rectangle, minimum width=1ex,rounded corners,draw] {\normalcolor\m@th$\displaystyle#1$};}}
 \makeatother
\newcommand*{\boxcolorg}{green}
\makeatletter
\newcommand{\boxedg}[1]{\textcolor{\boxcolorg}{%
\tikz[baseline={([yshift=-1ex]current bounding box.center)}] \node [rectangle, minimum width=1ex,rounded corners,draw] {\normalcolor\m@th$\displaystyle#1$};}}
 \makeatother
 \newcommand*{\boxcolorp}{purple}
\makeatletter
\newcommand{\boxedp}[1]{\textcolor{\boxcolorp}{%
\tikz[baseline={([yshift=-1ex]current bounding box.center)}] \node [rectangle, minimum width=1ex,rounded corners,draw] {\normalcolor\m@th$\displaystyle#1$};}}
 \makeatother
  \newcommand*{\boxcolorc}{cyan}
\makeatletter
\newcommand{\boxedc}[1]{\textcolor{\boxcolorc}{%
\tikz[baseline={([yshift=-1ex]current bounding box.center)}] \node [rectangle, minimum width=1ex,rounded corners,draw] {\normalcolor\m@th$\displaystyle#1$};}}
 \makeatother
  \newcommand*{\boxcolory}{yellow}
\makeatletter
\newcommand{\boxedy}[1]{\textcolor{\boxcolory}{%
\tikz[baseline={([yshift=-1ex]current bounding box.center)}] \node [rectangle, minimum width=1ex,rounded corners,draw] {\normalcolor\m@th$\displaystyle#1$};}}
 \makeatother

\usepackage{amsmath}

\title{Coulomb Branch Operator Algebras and \\[3mm] Universal~Selection~Rules~for~$\CN=2$~SCFTs}

\author{Matthew Buican}

\affiliation{\smallskip CTP and Department of Physics and Astronomy\\
Queen Mary University of London, London E1 4NS, UK}

\abstract{Coulomb branches of vacua are the most universal moduli spaces that arise in local unitary interacting 4d $\CN=2$ superconformal field theories (SCFTs). In these theories, $1/2$-BPS primaries parameterize the Coulomb branches and form (anti-)chiral rings. We define the notion of a Coulomb branch operator algebra, $\CA_{\CC}$, that contains these chiral and anti-chiral rings along with infinitely many more operators and products that are less protected by supersymmetry. Using a universal symmetry, $\CI\cong\mathbb{Z}_2$, that arises from studying the superconformal group, we give $\CI$ selection rules for $\CA_{\CC}$ and, more generally, for arbitrary products in the local operator algebra of any 4d $\CN=2$ SCFT. Defining the notion of a \lq\lq Coulombic" SCFT, we propose explanations for certain phenomena in a 4d/2d correspondence involving 4d $\CN=2$ theories and 2d vertex operator algebras. Finally, by considering deformations of $\CI$, we explore the case of $\CN>2$ SCFTs.}

\begin{document}
\setcounter{tocdepth}{2}
\maketitle
\toc

\newsec{Introduction}
In order to \lq\lq solve" a conformal field theory (CFT), we must construct an infinite number of correlation functions. Using consistency conditions from conformal symmetry, we can bootstrap the full set of correlation functions when there are a finite number of conformal primaries, as in the case of 2d minimal models \cite{Belavin:1984vu} (although there is at least one interesting case with infinitely many primaries that can also be bootstrapped \cite{Guillarmou:2020wbo}). In higher dimensions, we do not expect to find theories with a finite number of primaries, but it may still prove fruitful to focus on $n$-point functions of a small number of \lq\lq important" primaries (for ideas in this direction, see \cite{Rosenhaus:2018zqn,Bercini:2020msp,Poland:2023vpn}). This latter statement is particularly true when, as in free theories with finite central charge, there are a finite number of primaries that generate the operator algebra (see also the discussion in \cite{Harlow:2018tng}).

Thinking along these lines, it is interesting to note that, in superconformal field theories (SCFTs), there are special protected operator multiplets that are associated with renormalization group (RG) flows onto supersymmetric moduli spaces of vacua. One might hope that these multiplets play a special role in solving these theories.

For example, any local unitary interacting 4d $\CN=2$ SCFT is believed to have a Coulomb branch of vacua, $\CM_C$. General arguments (and all Lagrangian examples) suggest that there should be corresponding chiral and anti-chiral $1/2$-BPS primaries, $\CO_{\CC,i}$ and $\overline\CO_{\CC,i}$, sitting in short multiplets of type $\overline\CE_{r_i}$ and $\CE_{-r_i}$ generating chiral and anti-chiral rings, $\CR_{\CC}$ and $\overline\CR_{\CC}$ respectively, that give rise to coordinates on $\CM_{\CC}$ (note that for us these rings are operators in the UV SCFT).\footnote{Here we define $\bar\CE_{r_i}:=\bar\CE_{r_i(0,0)}$ in the multiplet notation of \cite{Dolan:2002zh} (and similarly for the conjugate multiplets). In the language of \cite{Cordova:2016emh}, $\bar\CE_{r_i}:=L\bar B_1[0;0]^{(0;2r_i)}_{r_i}$ and $\CE_{-r_i}:=B_1\bar L[0;0]^{(0;-2r_i)}_{r_i}$.} In particular, we can explore $\CM_C$ by turning on expectation values in the ultraviolet (UV) SCFT
\begin{equation}
\langle\CO_{\CC,i}\rangle~,\ \langle\overline\CO_{\CC,i}\rangle\ne0~,
\end{equation}
and flowing to the infrared (IR). One simple check of this picture is that, on the Coulomb branch, the superconformal $\mathfrak{u}(1)_R<\mathfrak{su}(2,2|2)$ should be spontaneously broken while $\mathfrak{su}(2)_R<\mathfrak{su}(2,2|2)$ should not. Indeed, the $\CO_{\CC,i}$ and $\overline\CO_{\CC,i}$ are charged under $\mathfrak{u}(1)_R$ (with respective charges $r_i,-r_i\ne0$) but are neutral under $\mathfrak{su}(2)_R$.

Moreover, the $\CO_{\CC,i}$ operators (and their complex conjugates) play an important role in the Seiberg-Witten construction of 4d $\CN=2$ gauge theories \cite{Seiberg:1994rs,Seiberg:1994aj}: their expectation values give rise to curves fibered over the resulting Coulomb branches that, in principle, allow one to exactly solve the two-derivative low-energy theories.

Given the significance of multiplets related to the Coulomb branch, a central question motivating this note is the following: what can the operator algebra, $\CA_{\CC}$, of the $\CO_{\CC,i}$, $\overline\CO_{\CC,i}$, and their descendants tell us about the UV SCFT at the origin of $\CM_{\CC}$ (i.e., the point at which $\langle\CO_i\rangle=\langle\overline\CO_i\rangle=0$ for all $i$)?

The most basic question to ask in this regard is: which multiplets can sit in $\CA_{\CC}$? To fruitfully rephrase this question, let us define $\CA_{\CC}$ as the following space (equipped with the OPE)
\begin{equation}\label{AC}
\CA_{\CC}:={\rm Span}\Big\{\Upsilon_b\in\CH_{L}:\Upsilon_b\in\bar\CE_{r_1}^{\times n_1}(x_{1,a_1})\times\cdots\times\bar\CE_{r_N}^{\times n_N}(x_{N,a_N})\times\CE_{-r_1}^{\times \bar n_1}(y_{1,a_1})\times\cdots\times\CE_{-r_N}^{\times \bar n_{N}}(y_{N,a_{N}})\Big\}~,
\end{equation}
where $\CH_{L}$ is the space of local operators,\footnote{Here and throughout this paper we only discuss {\it genuine} local operators (e.g., we do not discuss operators that live at the end of Wilson lines).} and $\Upsilon_b$ is a multiplet of the 4d $\CN=2$ superconformal algebra that sits in some (higher) OPE of $\CE$ and / or $\bar\CE$ multiplets.\footnote{Note that, in taking the span in \eqref{AC}, we allow for linear combinations involving different component operators appearing in the $\Upsilon_b$ multiplets. \label{Spanfoot}} The above product involves all operators in the corresponding generating $\bar\CE$ and $\CE$ multiplets (and derivatives of these operators). For finite central charge, $N$ is finite (moreover, in all known theories, $N$ corresponds to the \lq\lq rank" or complex dimension of $\CM_C$). In writing \eqref{AC}, we have defined the following (higher) OPE (note that this OPE is {\it not} the chiral ring product)
\begin{equation}\label{ChirOPEDef}
\bar\CE_{r_i}^{\times n_i}(x_{i,a_i}):=\bar\CE_{r_i}(x_{i,1})\times\bar\CE_{r_i}(x_{i,2})\times\cdots\times\bar\CE_{r_i}(x_{i,n_i})~,
\end{equation}
and similarly for the $\CE^{\times \bar n_i}_{-r_i}$ factors.\footnote{In \eqref{ChirOPEDef}, we understand the product $\bar\CE_{r_i}(x_{i,1})\times\bar\CE_{r_i}(x_{i,2})$ as shorthand for the OPEs between all components of the $\bar\CE_{r_i}$ multiplet (and all their conformal descendants). When we write $\Upsilon_b\in\bar\CE_{r_i}(x_{i,1})\times\bar\CE_{r_i}(x_{i,2})$, we mean that elements of the $\Upsilon_b$ multiplet are contained in the component OPEs. Then, $(\bar\CE_{r_i}(x_{i,1})\times\bar\CE_{r_i}(x_{i,2}))\times \bar\CE_{r_i}(x_{i,3})$ should be understood as the OPEs of all elements of the $\bar\CE_{r_i}(x_{i,3})$ multiplet (and their descendants) with all the operators appearing in $\bar\CE_{r_i}(x_{i,1})\times\bar\CE_{r_i}(x_{i,2})$. We define all higher OPEs recursively. Similar comments apply to the more general OPEs that include $\CE_{-r_i}$ in \eqref{AC}. All of these OPEs can be conveniently written in terms of superspace (although the details of such expressions will not be important for us below).}  Our question then becomes one of finding the most general set of $\Upsilon_b$ multiplets.\footnote{We emphasize that this question is much more general (and difficult) than finding the chiral and anti-chiral rings, since these latter structures sit properly inside $\CA_{\CC}$ via normal-ordered products
\begin{equation}
\CR_{\CC}~,\ \overline\CR_{\CC}<\CA_{\CC}~.
\end{equation}
To understand this point, note that the energy-momentum tensor satisfies
\begin{equation}\label{Tincl}
T_{\mu\nu}\in\hat\CC_{0(0,0)}\in\bar\CE_{r_1}\times\CE_{-r_1}\in\CA_{\CC}~,\ \ \ \hat\CC_{0(0,0)}\not\in\CR_{\CC}~,\ \overline\CR_{\CC}~,
\end{equation}
where $\hat\CC_{0(0,0)}$ is the energy-momentum tensor multiplet in the notation of \cite{Dolan:2002zh} (in the notation of \cite{Cordova:2016emh}, this is an $A_2\bar A_2[0;0]_2^{(0;0)}$ multiplet). Note that $\CA_{\CC}$ also controls infinitely many protected and unprotected contributions to arbitrary $n$-point functions of $\bar\CE$ and $\CE$ multiplets. \label{ACRC}
} 

Some intuition in this regard comes from thinking about Lagrangian theories with enough matter (e.g., $SU(N)$ $\CN=2$ SQCD with $N_f=2N$ flavors). Such SCFTs have, in addition to Coulomb branches, Higgs branches of vacua, $\CM_{\CH}$. These are moduli spaces of vacua where matter operators (e.g., mesons and baryons), transforming as primaries of $\hat\CB_{R}$ representations (or $B_1\bar B_1[0;0]_{2R}^{(2R;0)}$ multiplets in the language of \cite{Cordova:2016emh}) have non-zero vevs. The highest $\mathfrak{su}(2)_R$-weight primaries in these multiplets also form a chiral ring, $\CR_{\CH}$, called the Higgs branch chiral ring (as in the case of $\CR_{\CC}$, we define these operators to live in the UV SCFT {\it not} on the IR moduli space),\footnote{The lowest $\mathfrak{su}(2)_R$-weight primaries in these multiplets form the Higgs branch anti-chiral ring, $\overline\CR_{H}$.} that conjecturally satisfies \cite{Beem:2017ooy} 
\begin{equation}
\CM_{\CH}={\rm Spec}(\CR_{\CH})~,\ \ \ \CR_{\CH}=\mathbb{C}[\CM_{\CH}]~.
\end{equation}
The $\hat\CB_R$ primaries transform under $\mathfrak{su}(2)_R$ and are neutral under $\mathfrak{u}(1)_R$ (the reverse of the situation with Coulomb branch multiplets). Moreover, they are often charged under continuous flavor symmetries, while the $\bar\CE_{r_i}$ and $\CE_{-r_i}$ multiplets are always continuous flavor-neutral. Therefore, away from the origin of $\CM_H$, continuous flavor and $\mathfrak{su}(2)_R$ are spontaneously broken while $\mathfrak{u}(1)_R$ is not.

One consequence of the above discussion is that the intersections of the Higgs branch and Coulomb branch rings of a UV SCFT contain only the identity operator: $\CR_H\cap\CR_{\CC},\CR_H\cap\overline\CR_{\CC}=\mathds{1}$. Indeed, this statement follows from the $\mathfrak{su}(2)_R\oplus \mathfrak{u}(1)_R$ quantum numbers of the operators involved and is not limited to Lagrangian theories. Another consequence is that, when a $\hat\CB_R$ multiplet is charged under a continuous flavor symmetry, it cannot be in $\CA_{\CC}$. This statement also holds beyond the class of Lagrangian theories and follows from the $\bar\CE_{r}\times\CE_{-r}$ OPE \cite{Buican:2013ica,Buican:2014qla}.

Naively, this discussion suggests that $\hat\CB_R\not\in\CA_{\CC}$. But this argument also has glaring gaps. For example, we know that $\CA_{\CC}$ is always strictly larger than $\CR_{\CC}\cup\overline\CR_{\CC}$ (see footnote \ref{ACRC}), and we also know there are theories that have Higgs branches but no continuous flavor symmetry (e.g., see \cite{Hanany:2010qu}).\footnote{More generally, it is conceptually possible for there to exist continuous flavor-neutral $\hat\CB_R$ multiplets that are not related to moduli spaces.} In fact, we will see later that superconformal selection rules on their own do not forbid certain $\hat\CB_R$ multiplets from appearing in $\CA_{\CC}$.

Therefore, to get a more universal handle on $\CA_{\CC}$ and on any other operator subalgebras that might appear in unitary local 4d $\CN=2$ SCFTs, it is essential to revisit the superconformal group itself. One useful perspective is to note that the 4d $\CN=2$ superconformal symmetry is, as an algebra, $\mathfrak{su}(2,2|2)$, but that, as a group, it can take different forms. In particular, let us consider the $\mathfrak{su}(2)_R\oplus\mathfrak{so}(4)<\mathfrak{su}(2,2|2)$ subalgebra and the corresponding simply connected group, $SU(2)_R\times Spin(4)$.\footnote{We work in Euclidean space, but we can make similar comments in Lorentzian signature.} We can then study the action of
\begin{equation}
\CI:\cong{\rm Diag}(Z(SU(2)_R)\times Z(Spin(4)))\cong\mathbb{Z}_2~,
\end{equation}
on local operators. In fact, since $\CI$ is guaranteed to commute with the rest of the superconformal group, $\CI$ can also be thought of as a flavor symmetry (this perspective has been emphasized in \cite{Distler:2020tub,Heckman:2022suy}) and so we can study the action of $\CI$ multiplet-by-multiplet (i.e., each operator in the same multiplet has the same $\CI$ charge, since the supercharges are $\CI$ neutral). As we will illustrate in the main text through many examples and different classes of theories, $\CI$ may be trivial in some theories and non-trivial in others.\footnote{In an interesting recent pair of papers \cite{Cordova:2018acb,Brennan:2022tyl}, the authors studied $\CN=2$ $SU(2)$ SYM and $\CN=2$ $SU(2)$ SQCD as well as certain non-supersymmetric cousins. These $\CN=2$ theories have trivial $\CI$ action on gauge-invariant local operators (though, in the case of $\CN=2$ $SU(2)$ SQCD, not on the fundamental hypermultiplets). This fact is related to the ability to put these theories on non-spin manifolds (see also \cite{Aspman:2022sfj} for a discussion from the point of view of topological twisting) and to detect various interesting 't Hooft anomalies. We will briefly comment on these links in the discussion section at the end of this paper. We thank A.~Antinucci, P.~Genolini, and L.~Tizzano for emphasizing these points to us.}

As we will explain in Sec. \ref{VOAspins}, understanding when $\CI$ is trivial potentially leads to a deeper understanding of certain phenomena in 4d $\CN=2$ SCFTs that have, until now, appeared to be coincidental. 

To set up these results, we begin by noting the trivial fact that any multiplet of local operators in a 4d $\CN=2$ SCFT with definite superconformal quantum numbers, $\Upsilon_a$, lives in one and only one of the following two sectors
\begin{equation}\label{PMsectors}
\CA_{\CO,\pm}:={\rm Span}\left\{\Upsilon_a\in\CH_L: \CI(\Upsilon_a)=\pm1\right\}~,
\end{equation}
where \lq\lq $\CI(\cdots)$" denotes the charge of the enclosed multiplet (comments similar to those in footnote \ref{Spanfoot} apply for linear combinations of operators in $\CA_{\CO,\pm}$). Moreover, $\CA_{\CO,+}$ forms a closed operator subalgebra and must be non-trivial in any local 4d $\CN=2$ SCFT since
\begin{equation}
T_{\mu\nu}\in\hat\CC_{0(0,0)}\in\CA_{\CO,+}~.
\end{equation}
On the other hand, $\CA_{\CO,-}$ cannot be closed under the OPE and may or may not be trivial.

More generally, this discussion immediately leads to selection rules in all $n$-fold OPEs of any 4d $\CN=2$ SCFT that we will explore in more detail below. For example, we will study the constraints the selection rules place on relations between the Coulomb branch algebra, $\CA_{\CC}$, the Higgs branch operators, and certain vertex operator algebras (VOAs) that are associated with any local 4d $\CN=2$ SCFT \cite{Beem:2013sza}.

The plan of this paper is as follows. In the next section, we introduce the set of allowed unitary irreducible representations of $\mathfrak{su}(2,2|2)$, their charges under $\CI$, and the resulting OPE selection rules. We emphasize the connection between $\CI$ and properties of the associated 2d VOAs alluded to above. Then, in Sec. \ref{ACgen}, we move on to discuss constraints on the operators that can appear in the $\CA_{\CC}$ algebra. For example, we comment on when $\hat\CB_R$ Higgs branch multiplets are forbidden from appearing and when they are allowed to appear (at the level of superconformal selection rules). In Sec. \ref{Coulombic}, we prove some theorems on 4d $\CN=2$ SCFTs whose operator algebra is generated by the Coulomb branch $\bar\CE_r$ and $\CE_{-r}$ multiplets (we call these theories \lq\lq Coulombic"). Next, in Sec. \ref{VOAspins} we use one of these theorems to give a potential explanation for why certain 2d VOAs associated with 4d $\CN=2$ SCFTs have only integral 2d spins. We then generalize our discussion to the larger superconformal algebras with $\CN>2$ SUSY and explain some general constraints related 2d VOAs must satisfy. Finally, we conclude with a discussion of open problems.\footnote{In this paper we do not solve the problem of giving the full set of constraints that superconformal selection rules put on $\CA_{\CC}$. Additional progress on this problem will be reported elsewhere, in collaboration with C.~Bhargava, H.~Jiang, and T.~Nishinaka \cite{2024paper}. Instead, here we focus on the constraints arising purely from $\CI$.}

\newsec{4d $\CN=2$ irreps, OPEs, and the $\CI$ selection rules}
We will primarily think of unitary local 4d $\CN=2$ SCFTs from an abstract algebraic perspective, because our main interest will be to study model-independent selection rules for the OPEs and correlation functions of multiplets of the corresponding superconformal algebra, $\mathfrak{su}(2,2|2)$. A logical starting point for this discussion is the list of unitary irreducible representations (up to CPT conjugates, like $\CE_{-r}$, that we drop for simplicity) \cite{Dolan:2002zh}\footnote{Alternatively, in the somewhat more elaborate notation of \cite{Cordova:2016emh}, we have (note that for us, $r_{\rm here} = {1\over2}r_{\rm there}$, $j_{\rm here} = {1\over2}j_{\rm there}$, and $\bar j_{\rm here} = {1\over2}\bar j_{\rm there}$ relative to the conventions in \cite{Cordova:2016emh})
\begin{eqnarray}\label{4dN2irrepsC}
&&L\bar B_1[0;0]_{r}^{(0;2r)}~,\ B_1\bar B_1[0;0]_{2R}^{(2R;0)}~,\ A_1\bar B_1[2j\ge1;0]_{2R+r}^{(2R;2r=2j+2)}~, A_2\bar B_1[0;0]^{2R; 2r=2}~,\ L\bar B_1[2j\ge1;0]_{2R+r}^{(2R;2r>2j+2)}~,\ \cr&& A_1\bar A_1[2j\ge1;2\bar j\ge1]^{2R;2r=2j-2\bar j}_{2+2R+2j-2\bar j}~,\ A_1\bar A_2[2j\ge1;0]_{2+2R+j}^{(R;2r=2j)}~,\ A_2\bar A_2[0;0]_{2+2R}^{(R;0)}~,\ L\bar A_1[2j;2\bar j\ge1]_{2+2R+2\bar j+r}^{(R;2r>2j-2\bar j)}~,\ \cr &&L\bar A_2[2j;0]_{2+2R+r}^{(2R;2r>2j)}~,\ L\bar L[2j;2\bar j]_{\Delta}^{(2R;2r)}
\end{eqnarray}
In the above list, the first entry corresponds to $\bar\CE_r$, the second entry corresponds to $\hat\CB_R$, $A_1\bar B_1$ and $A_2\bar B_1$ correspond to $\bar\CD$ ($A_2\bar B_1$ has a scalar primary and hence a different set of shortening conditions), the fifth entry corresponds to $\bar\CB_{R,r(j,0)}$, $A_1\bar A_1$, $A_1\bar A_2$, and $A_2\bar A_2$ correspond to $\hat\CC_{R(j,\bar j)}$ (with different values of the spin and hence different shortening conditions), and $L\bar A_1$ and $L\bar A_2$ correspond to $\bar\CC_{R,r(j,\bar j)}$ (with different values of the spin and therefore different shortening conditions).
}
\begin{equation}\label{4dN2irreps}
\bar\CE_r~,\ \hat\CB_R~,\ \bar\CD_{R(j,0)} ~,\ \bar\CB_{R,r(j,0)}~,\ \hat\CC_{R(j,\bar j)}~,\ \bar\CC_{R,r(j,\bar j)}~,\ \CA^{\Delta}_{R,r(j,\bar j)}~,
\end{equation}
where $r$ corresponds to the superconformal $\mathfrak{u}(1)_R$ charge of the primary, $R$ to the $\mathfrak{su}(2)_R$ spin of the primary, while $j$ and $\bar j$ correspond to the $\mathfrak{so}(4)$ spins, and $\Delta$ is the scaling dimension.\footnote{Note that $\Delta$ does not appear explicitly in the first six multiplets of \eqref{4dN2irreps} because, as short multiplets, their scaling dimensions are fixed in terms of the other quantum numbers.}

Let us briefly discuss some of the properties of these multiplets (referring the interested reader to \cite{Dolan:2002zh} for a full account of the technical details):
\begin{itemize}
\item The $\bar\CE_r$ multiplets have $R=j=\bar j=0$, $\Delta=r>1$, and correspond to generalizations of the Casimirs---${\rm Tr}\Phi^n$---in Lagrangian theories (where $\Phi$ is the adjoint-valued chiral gauge multiplet). In all known theories, vevs of the $\bar\CE_r$ primaries parameterize the Coulomb branch and form the ring $\CR_{\CC}$ described in the introduction. The primaries are annihilated by all anti-chiral supercharges, $\bar Q^i_{\dot\alpha}$ (where $i=1,2$ is an $\mathfrak{su}(2)_R$ index, and $\dot\alpha$ is a right spinor index of $\mathfrak{su}(2)_-<\mathfrak{so}(4)$). Strictly speaking, the list of unitary irreducible representations of $\mathfrak{su}(2,2|2)$ contains multiplets with spinning primaries, $\bar\CE_{r(j,0)}$ \cite{Dolan:2002zh}, but locality implies that $j=0$ \cite{Manenti:2019jds}. Therefore, since we are interested in local theories, we ignore such representations and define $\bar\CE_r:=\bar\CE_{r(0,0)}$. As mentioned in the introduction, along with the conjugate $\CE_{-r}$ multiplets, the $\bar\CE_r$ irreps generate the $\CA_{\CC}$ algebra described around \eqref{AC}.
\item The $\hat\CB_R$ multiplets have $R\ge1/2$, $j=\bar j=0$, $\Delta=2R$, and have highest $\mathfrak{su}(2)_R$ components corresponding to generalizations of the mesons and baryons---$q\tilde q$, $q^N$, $\tilde q^N$---in Lagrangian theories (where $q$ and $\tilde q$ correspond to the matter hypermultiplets in representation $R\oplus \overline{R}$ of the gauge group). The highest $\mathfrak{su}(2)_R$-weight primary is annihilated by the highest $\mathfrak{su}(2)_R$-weight chiral and anti-chiral supercharges. In all known theories, vevs of $\hat\CB_R$ primaries parameterize the Higgs branch, and highest $\mathfrak{su}(2)_R$ components form the ring, $\CR_{\CH}$, described in the introduction.
\item The $\bar\CD_{R(j,0)}$ multiplets have $R,j\ge0$ (locality implies $R\ge j$ \cite{Manenti:2019kbl,Buican:2023efi}), $r=1+j$, and $\Delta=2R+1+j$. Examples of such multiplets include $\bar\CD_{1/2(0,0)}$, which contain extra SUSY currents and are present in all local $\CN>2$ SCFTs. The highest $\mathfrak{su}(2)_R$-weight state of the primaries is annihilated by the highest $\mathfrak{su}(2)_R$-weight anti-chiral supercharge, $\bar Q^1_{\dot\alpha}$ (the multiplet also satisfies a semi-shortening condition with respect to the chiral supercharge). When $j=0$, $\bar\CD_{R(j,0)}$ multiplets can parameterize mixed branches.
\item The $\bar\CB_{R,r(j,0)}$ multiplets have have $R,j\ge0$ (locality is conjectured to imply $R\ge j$ \cite{Banerjee:2023ddh,Buican:2023efi}), $r>1+j$, and $\Delta=2R+r$. These multiplets have been studied extensively in \cite{Banerjee:2023ddh}, where they were shown to arise in various contexts, including from chiral ring products of $\bar\CE$ and $\hat\CB_R$ primaries when there is a mixed branch and also from products of $\bar\CE$ primaries and descendants (these statements will be useful for us below). The highest $\mathfrak{su}(2)_R$-weight primaries are annihilated by the highest $\mathfrak{su}(2)_R$-weight anti-chiral supercharge, $\bar Q^1_{\dot\alpha}$.
\item The $\hat\CC_{R(j,j)}$ multiplets have $R,j,\bar j\ge0$ (locality is conjectured to imply $R\ge|j-\bar j|-1$ \cite{Buican:2023efi}), $r=j-\bar j$, and $\Delta=2+2R+j+\bar j$. A prominent example of such a multiplet includes the energy-momentum tensor multiplet, $\hat\CC_{0(0,0)}$, which is present in any local 4d $\CN=2$ SCFT (in a Lagrangian theory, the primary takes the schematic form $J\sim{\rm Tr}(\Phi^{\dagger}\Phi)+\sum_{R}q_R^{\dagger}q_R$). These multiplets satisfy semi-shortening conditions with respect to the chiral and anti-chiral supercharges. In particular, such multiplets are non-chiral and must exist in any operator algebra that contains multiplets and their CPT conjugates, since $T_{\mu\nu}$ must appear in the corresponding OPEs. The previous sentence explains the discussion around \eqref{Tincl} in footnote \ref{ACRC}.
\item The $\bar\CC_{R,r(j,\bar j)}$ multiplets have $R,j,\bar j\ge0$ (locality is conjectured to imply $R\ge j-\bar j-1$ \cite{Buican:2023efi}), $r>j-\bar j$, and $\Delta=2+2R+2\bar j+r$. Important examples of such multiplets were discussed in \cite{Bhargava:2022cuf,Bhargava:2022yik}. Since these multiplets are the least protected short representations (they satisfy semi-shortening conditions with respect to the anti-chiral supercharges), they have the most to say about the \lq\lq global" form of the local operator algebra. For example, the results of \cite{Bhargava:2022yik} show that, in the minimal Argyres-Douglas (MAD) SCFT (using the terminology of \cite{Bhargava:2022cuf,Bhargava:2022yik} or, in other nomenclature, the $(A_1, A_2)$ SCFT), all $\bar\CC_{0,r(0,0)}$ multiplets arise from the $\bar\CE_r\times\hat\CC_{0(0,0)}$ OPE, and all $\bar\CC_{0,n(1,0)}$ multiplets arise from the $\bar\CE_r\times\bar\CE_r$ OPE. This discussion suggests that the MAD theory has a local operator algebra generated by $\bar\CE_r$ and $\CE_{-r}$ multiplets (we will examine some consequences of this suggestion below).
\item $\CA^{\Delta}_{R,r(j,\bar j)}$ is a long multiplet with $R,j,\bar j\ge0$, with $r$ and $\Delta$ unfixed (although unitarity implies $\Delta>2+2R+{\rm max}\left\{2j-r,2\bar j+r\right\}$). Such a multiplet does not satisfy any shortening conditions, but it does break up into a collection of short multiplets when $\Delta$ hits the unitarity bound.
\end{itemize}

Given these representations, we would like to study the resulting OPEs and correlation functions. By superconformal symmetry, we can reduce any such $n$-point function through the multiplet three-point functions
\begin{equation}\label{3pt}
\langle\Upsilon_{\CI_1}(z_1)\Upsilon_{\CI_2}(z_2)\Upsilon_{\CI_3}(z_3)\rangle={T_{\CI_1}^{\CJ_1}T^{\CJ_2}_{\CI_2}\over(x^2_{1\bar3})^{q_1}(x^2_{1\bar3})^{\bar q_1}(x^2_{2\bar3})^{{q}_2}(x^2_{\bar23})^{\bar q_2}}H_{\CJ_1\CJ_2\CI_3}({\bf X}_{3\alpha\dot\alpha}, \Theta_3^{i\alpha},\bar\Theta^{\dot\alpha}_{3i})~.
\end{equation}
The precise technical details of this expression are rather involved and will not be important for us in what follows (instead, we refer the interested reader to \cite{Park:1999pd,Kuzenko:1999pi,Nirschl:2004pa,Kiyoshige:2018wol} for a full discussion).\footnote{Here for completeness we merely comment that, in writing \eqref{3pt}, we have defined $\CI_a, \CJ_a$ as combined $\mathfrak{su}(2)_R\oplus\mathfrak{so}(4)$ indices, $i$ is an $\mathfrak{su}(2)_R$ index, $(\alpha,\dot\alpha)$ are $\mathfrak{so}(4)$ indices, the $T_{\CI_a}^{\CJ_a}$ are tensors built from the superspace coordinates, $X_3$ is Grassmann-even, $\Theta_3$ and $\bar\Theta_3$ are Grassmann odd, the $x_{i\bar j}$ and $x_{\bar i j}$ are Grassmann even, and the $q_i$'s and $\bar q_i$'s are related to the quantum numbers of the $\Upsilon$ multiplets. Here Grassmann evenness / oddness is measured with respect to the $\theta_{a,\alpha}^i$'s and $\bar\theta_{a\dot\alpha}^i$'s (where $a=1,2,3$ label the Grassmann coordinates associated with the three superspace points in the correlation function).\label{Comp}} By imposing shortening conditions on the $\Upsilon_{\CI_a}$, we generate a set of selection rules that superconformal representation theory imposes on the OPE, and we can determine if it is consistent for $\overline\Upsilon_{\overline\CI_3}$ to appear in the OPE of the first two superfields in \eqref{3pt}
\begin{equation}
\Upsilon_{\CI_1}(z_1)\times\Upsilon_{\CI_2}(z_2)\ni\overline\Upsilon_{\overline\CI_3}~.
\end{equation}
Of course, the question of whether  the OPE coefficient for an allowed $\Upsilon$ vanishes or not is typically a question of dynamics or of symmetries beyond the superconformal algebra.

In general, it is rather involved to derive even the superconformal selection rules.\footnote{Although in certain cases, these rules have been derived and put to great effect as in \cite{Liendo:2015ofa}. For certain shortcuts in the case of non-singular terms in the OPE, see \cite{Agarwal:2018zqi}. In an upcoming paper, we will study some of the superconformal selection rules for $\CA_{\CC}$ in more detail \cite{2024paper}.} One well-known reason is that the OPEs of superconformal primaries do not determine the OPEs of all superconformal descendants. However, as we have alluded to above, additional symmetries can help. For example, following \cite{Buican:2013ica,Buican:2014qla}, supersymmetric Ward identities imply that any chiral operator, $\CO$, charged under a flavor Lie algebra, $\mathfrak{g}$, must satisfy
\begin{equation}\label{ChirOPE}
\bar\CO(x_1)\times\CO(x_2)\ni J_{\mathfrak{g}}~,
\end{equation}
where $J_{\mathfrak{g}}$ is the $\mathfrak{su}(2)_R$-weight zero component of the $\mathfrak{su}(2)_R$ spin-one primary of a $\hat\CB_1$ multiplet (such multiplets necessarily contain flavor symmetry Noether currents). Taking $\CO$ to be a primary of a $\bar\CE_r$ multiplet, we see that it is $\mathfrak{su}(2)_R$-neutral and so \eqref{ChirOPE} is inconsistent. This result implies that the whole $\bar\CE_r$ multiplet must be neutral under $\mathfrak{g}$. Therefore, we see that if a multiplet $\Upsilon$ transforms under some non-trivial $\mathfrak{g}$ irrep, $R_{\mathfrak{g}}$, we must have
\begin{equation}
R_{\mathfrak{g}}(\Upsilon)\ne1~\ \ \ \Rightarrow\ \ \ \Upsilon\not\in\CA_{\CC}~,
\end{equation}
where we defined $\CA_{\CC}$ in \eqref{AC} and repeat the definition again below for convenience 
\begin{equation}\label{AC2}
\CA_{\CC}:={\rm Span}\Big\{\Upsilon_b\in\CH_{L}:\Upsilon_b\in\bar\CE_{r_1}^{\times n_1}(x_{1,a_1})\times\cdots\times\bar\CE_{r_N}^{\times n_N}(x_{N,a_N})\times\CE_{-r_1}^{\times \bar n_1}(y_{1,a_1})\times\cdots\times\CE_{-r_N}^{\times \bar n_{\bar N}}(y_{\bar N,a_{\bar N}})\Big\}~.
\end{equation}
However, as described in the introduction, there are many 4d $\CN=2$ SCFTs that do not have Lie group flavor symmetries, and the above analysis has nothing to say about these cases.

Therefore, we are motivated to consider the $\CI$ symmetry discussed in the introduction
\begin{equation}\label{DefI}
\CI:\cong{\rm Diag}(Z(SU(2)_R)\times Z(Spin(4)))\cong\mathbb{Z}_2~.
\end{equation}
At the level of the superconformal quantum numbers introduced above, we can take the generator of $\CI$ to be
\begin{equation}
\rho_{\CI}:=e^{2\pi i(R+j+\bar j)}~.
\end{equation}
This generator leaves all supercharges invariant since they transform as $(1/2,1/2,0)$ and $(1/2,0,1/2)$ irreps under $\mathfrak{su}(2)_R\oplus\mathfrak{so}(4)$. It clearly also leaves the bosonic part of $\mathfrak{su}(2,2|2)$ invariant and so $\rho_{\CI}$ is in the center of the superconformal group. 

Given the above definition, it is easy to see that the superconformal representations split in the following way under the action of $\CI$ (as in \eqref{PMsectors}, the \lq\lq$\pm$" superscript refers to the $\CI$ charge; note that we also drop CPT-conjugate multiplets below)
\begin{eqnarray}\label{Apm}
\CA_{\CO,+}&\ni&\bar\CE_r~,\ \hat\CB_R|_{R\in\mathbb{Z}}~,\ \bar\CD_{R(j,0)}|_{R+j\in\mathbb{Z}} ~,\ \bar\CB_{R,r(j,0)}|_{R+j\in\mathbb{Z}}~,\ \hat\CC_{R(j,\bar j)}|_{R+j+\bar j\in\mathbb{Z}}~,\ \bar\CC_{R,r(j,\bar j)}|_{R+j+\bar j\in\mathbb{Z}}\cr&& \ \  \CA^{\Delta}_{R,r(j,\bar j)}|_{R+j+\bar j\in\mathbb{Z}}~, \cr \CA_{\CO,-}&\ni&\hat\CB_R|_{R\in{1\over2}\mathbb{Z}_{\rm odd}}~,\ \bar\CD_{R(j,0)}|_{R+j\in{1\over2}\mathbb{Z}_{\rm odd}} ~,\ \bar\CB_{R,r(j,0)}|_{R+j\in{1\over2}\mathbb{Z}_{\rm odd}}~,\ \hat\CC_{R(j,\bar j)}|_{R+j+\bar j\in{1\over2}\mathbb{Z}_{\rm odd}}~,\cr&& \ \ \bar\CC_{R,r(j,\bar j)}|_{R+j+\bar j\in{1\over2}\mathbb{Z}_{\rm odd}}~,\ \CA^{\Delta}_{R,r(j,\bar j)}|_{R+j+\bar j\in{1\over2}\mathbb{Z}_{\rm odd}}~.\ \ \ \ \ \ \ 
\end{eqnarray}
This discussion also implies the following simple but important fact, which we will state as a theorem:

\bigskip
\noindent
{\bf Theorem 1:}  Any 4d $\CN=2$ SCFT respects an $\CI$ selection rule. In particular, for any $\mathfrak{su}(2,2|2)$ irreducible representations $\Upsilon$, $\Upsilon_1$,$\cdots$, $\Upsilon_m$, we have
\begin{equation}\label{SR}
\Upsilon\in\Upsilon_1^{\times n_1}\times\Upsilon_2^{\times n_1}\times\cdots\times\Upsilon^{\times n_m}_m\ \Rightarrow\ \CI(\Upsilon)=\CI(\Upsilon_1)^{n_1}\cdots\CI(\Upsilon_m)^{n_m}~.
\end{equation}

\medskip
\noindent
We will return to this selection rule at various points in the discussion below.

\subsec{Selection rules, the Schur sector, and the associated VOA}\label{SchurSec}
The discussion in the previous subsection interacts in interesting ways with the so-called \lq\lq Schur" sector of 4d $\CN=2$ operators and the associated 2d VOA \cite{Beem:2013sza}. We briefly review this sector of operators along with the 2d VOA and then explain a connection with the $\CI$ symmetry (for a fuller accounting of the Schur/VOA correspondence, we refer the reader to \cite{Beem:2013sza}).

To that end, recall that the Schur operators are highest $\mathfrak{so}(4)$-weight components of the highest $\mathfrak{su}(2)_R$-weight operators sitting in the following multiplets (for simplicity, as in \eqref{4dN2irreps}, we drop the CPT-conjugate $\CD_{R(0,j)}$ multiplet)
\begin{equation}
\hat\CB_R~,\ \bar\CD_{R(j,0)}~, \hat\CC_{R(j,\bar j)}~.
\end{equation}
In particular, the corresponding Schur multiplets sit in both $\CA_{\CO,\pm}$ sectors
\begin{eqnarray}
\CA_{\CO,+}>\CA_{\CO,+}^{\rm Schur}&\ni&\hat\CB_R|_{R\in\mathbb{Z}}~,\ \bar\CD_{R(j,0)}|_{R+j\in\mathbb{Z}} ~,\ \hat\CC_{R(j,\bar j)}|_{R+j+\bar j\in\mathbb{Z}}~, \cr \CA_{\CO,-}>\CA_{\CO,-}^{\rm Schur}&\ni&\hat\CB_R|_{R\in{1\over2}\mathbb{Z}_{\rm odd}}~,\ \bar\CD_{R(j,0)}|_{R+j\in{1\over2}\mathbb{Z}_{\rm odd}} ~,\  \hat\CC_{R(j,\bar j)}|_{R+j+\bar j\in{1\over2}\mathbb{Z}_{\rm odd}}~.
\end{eqnarray}
As we have already discussed in the introduction, $\CA_{\CO,-}$ is not closed under OPE. Therefore, neither is $\CA_{\CO,-}^{\rm Schur}$. Similarly, even though $\CA_{\CO,+}$ is closed under OPE, we expect that $\CA_{\CO,+}^{\rm Schur}<\CA_{\CO,+}$ is also never closed (otherwise, long multiplets and other non-Schur multiplets would not be allowed to appear in Schur-Schur OPEs).

More interestingly, we can rephrase the $\CI$ quantum number of any Schur operator, $\CO_S$, (and its multiplet) in terms of the spin of the corresponding 2d VOA current. Indeed, an important consequence of the 4d/2d map of \cite{Beem:2013sza} is that any Schur operator (really a certain $\mathfrak{su}(2)_R$-twisted-translated cohomology class, $[\CO_S]$) maps to a VOA current with 2d spin
\begin{equation}
h([\CO_s])=(R+j+\bar j)(\CO_S)~.
\end{equation}
Therefore, we arrive at the following theorem:

\bigskip
\noindent
{\bf Theorem 2:} Given any Schur operator, $\CO_S$, in a 4d $\CN=2$ SCFT
\begin{equation}\label{SchurCond}
\CO_{S}\in\CA^{\rm Schur}_{\CO,+}\ \Leftrightarrow\ h([\CO_{S}])\in\mathbb{Z}~,\ \ \ \CO_{S}\in\CA^{\rm Schur}_{\CO,-}\ \Leftrightarrow\ h([\CO_{S}])\in{1\over2}\mathbb{Z}_{\rm odd}~.
\end{equation}

\medskip
\noindent
In other words, the parity of the 2d spin of the 2d current related to a Schur operator is fixed by the $\CI$ charge of the 4d parent.\footnote{Note that $\CI$ does not fix the 2d (or 4d) statistics, since the VOA is necessarily non-unitary \cite{Beem:2013sza}.} 

For example, the $\hat\CC_{0(0,0)}$ energy-momentum tensor multiplet Schur operator is the highest $\mathfrak{su}(2)_R\oplus\mathfrak{so}(4)$-weight component of the $\mathfrak{su}(2)_R$ current, $J_{+\dot+}^{11}$. It is easy to check that $h=2$, which is consistent with its being mapped to the holomorphic stress tensor
\begin{equation}
[J_{+\dot+}^{11}]\cong T\ \ \Rightarrow\ h([J_{+\dot+}^{11}])=2~,\ \ \ \CI(\hat\CC_{0(0,0)})=1~.
\end{equation}

The simplest possible 2d VOAs are Virasoro minimal models. By the above discussion, any 4d ancestors of the corresponding currents have $\CI=1$, since the currents consist purely of integer-spin operators (normal ordered products of $T$ and its derivatives). Such VOAs are known to be realized in the simplest Argyres-Douglas $\CN=2$ SCFTs. Later on, we will suggest an explanation for these facts and other related phenomena in certain more general theories.

As another example, the extra supercurrent Schur multiplet, $\bar\CD_{1/2(0,0)}$, has a Schur operator corresponding to the highest $\mathfrak{su}(2)_R\oplus\mathfrak{so}(4)$-weight component among level-one descendants, $S^{11}_{+}$. The corresponding VOA state has $h=3/2$ and is consistent with
\begin{equation}
[S^{11}_{+}]\cong \bar G\ \ \Rightarrow\ h([S_{+}^{11}])=3/2~,\ \ \ \CI(\bar\CD_{1/2(0,0)})=-1~,
\end{equation}
where $\bar G$ is a 2d supercurrent. The CPT-conjugate $\CD_{1/2(0,0)}$ multiplet furnishes an additional supercurrent and, depending on whether $\CN=3,4$ in 4d, we have an $\CN=2,4$ 2d VOA with half-integer 2d spin currents. This discussion leads to a simple observation we will return to in Sec. \ref{Ng2}:

\bigskip
\noindent
{\bf Corollary 3:} Any local 4d SCFT with $\CN>2$ SUSY has non-trivial $\CI$ (i.e., $\CI$ acts faithfully on the local operator algebra of the theory).

\medskip
More generally, 4d $\CN=2$ SCFTs can (and often do) have trivial / unfaithful $\CI$. For example, the free vector multiplet has $\CI=1$ for any local operator (and the corresponding 2d VOA, the small algebra of a $(b,c)$ ghost system of weights $(1,0)$, has only integer 2d spins). Similar comments apply to $\CN=2$ $SU(2)$ SQCD with $N_f=4$. The reason is that any gauge-invariant operator will have an even number of hypermultiplets (these are $\hat\CB_{1/2}$ multiplets transforming in the doublet of $SU(2)$ and having $\CI=-1$) while the vector multiplet has $\CI=1$ (the corresponding VOA at generic gauge coupling, $\widehat{\mathfrak{so}(8)_{-2}}$, only has states of integer 2d spin).

\subsec{What can appear in the Coulomb branch operator algebra?}\label{ACgen}
Since all local unitary interacting 4d $\CN=2$ SCFTs are believed to have a Coulomb branch, it is natural to study the operator algebra for the corresponding chiral and anti-chiral UV $\CN=2$ multiplets, $\CA_{\CC}$. As mentioned at length in the introduction and previous sections, this algebra consists of the following operators equipped with the OPE
\begin{equation}\label{AC3}
\CA_{\CC}:={\rm Span}\Big\{\Upsilon_b\in\CH_{L}:\Upsilon_b\in\bar\CE_{r_1}^{\times n_1}(x_{1,a_1})\times\cdots\times\bar\CE_{r_N}^{\times n_N}(x_{N,a_N})\times\CE_{-r_1}^{\times \bar n_1}(y_{1,a_1})\times\cdots\times\CE_{-r_N}^{\times \bar n_{\bar N}}(y_{\bar N,a_{\bar N}})\Big\}~.
\end{equation}
Now, we saw in \eqref{Apm} that the $\bar\CE$ and $\CE$ multiplets have $\CI=+1$ and so all operators in $\CA_{\CC}$ must have $\CI=+1$
\begin{equation}
\CA_{\CC}\le\CA_{\CO,+}~,\ \ \ \CO\in\CA_{\CC}\ \Rightarrow\ (R+j+\bar j)(\CO)\in\mathbb{Z}~.
\end{equation}
Equivalently, we can say
\begin{equation}
\CA_{\CC}\cap\CA_{\CO,-}=\emptyset~,\ \ \  (R+j+\bar j)(\CO)\in{1\over2}\mathbb{Z}_{\rm odd}\ \Rightarrow \CO\not\in\CA_{\CC}~.
\end{equation}

For the Schur sector, combining this logic with theorem 2 shows that any Schur operator, $\CO_S$, and corresponding multiplet satisfies
\begin{equation}
\CO_S\in\CA_{\CC}\ \Rightarrow\ h([\CO_S])\in\mathbb{Z}~,\ \ \ h([\CO_s])\in{1\over2}\mathbb{Z}_{\rm odd}\ \Rightarrow\ \CO_S\not\in\CA_{\CC}~.
\end{equation}
In other words, a Schur multiplet appears in an $n$-fold OPE of $\bar\CE$ and / or $\CE$ multiplets only if the corresponding VOA state has integer 2d spin. Specializing to the Higgs branch multiplets, we then see that
\begin{equation}\label{BRcons}
R\in{1\over2}\mathbb{Z}_{\rm odd}\ \Rightarrow\ \hat\CB_R\not\in\CA_{\CC}~.
\end{equation}

This latter result is what we expect when we have a $\hat\CB_R$ multiplet that is charged under a continuous flavor symmetry (see the argument around \eqref{ChirOPE}). Examples include baryons in superconformal SQCD. However, in \eqref{BRcons} we see that $\hat\CB_R$ multiplets with half-integer $R$ are forbidden from appearing in $\CA_{\CC}$ even when no continuous flavor symmetry is present.\footnote{Note that the case of $\hat\CB_{1/2}$ (i.e., a free hyper) can be understood using \eqref{BRcons} or from the fact that such a multiplet necessarily has an $SU(2)$ flavor symmetry (of which we can identify $\CI$ with the center). Of course, dynamics also prevent such a contribution since, in order for a free hypermultiplet to couple to $\bar\CE$ or $\CE$ multiplets, we would need to have interactions.} 

What about the case of $\hat\CB_R$ multiplets with $R$ an integer that are uncharged under a continuous flavor symmetry (e.g., a $U(1)$ current multiplet, $\hat\CB_1$)?  We know these multiplets are also uncharged under $\CI$. From the fact that the Higgs branch and Coulomb branch chiral rings of the UV SCFT do not intersect (as discussed in the introduction, this statement follows from the fact that the latter is charged under $\mathfrak{u}(1)_R$ and the former is not), we might be tempted to conclude that $\hat\CB_{R\in\mathbb{Z}}$ multiplets cannot be in $\CA_{\CC}$. This statement may be true (we are not aware of any explicit counterexamples), but we would like to point out that superconformal selection rules themselves allow
\begin{equation}\label{BEEE}
\hat\CB_1^{\rm neutral}\in \bar\CE_{r_1}\times\bar\CE_{r_2}\times\CE_{-r_3}~, \ \ \ r_3=r_1+r_2-1~,
\end{equation}
where \lq\lq neutral" in the superscript on the lefthand side reminds us that the corresponding multiplet is neutral under continuous flavor symmetries (i.e., it is an Abelian current multiplet) and so, in principle, we can have
\begin{equation}\label{BRint}
\hat\CB_{R\in\mathbb{Z}}^{\rm neutral}\in\CA_{\CC}~.
\end{equation}
When this inclusion does not hold in particular theories, there must be some dynamical reason, some governing symmetry beyond those we are describing here, or some aspect of superconformal symmetry that goes beyond the selection rules.

One way to derive \eqref{BEEE} is to note that superconformal selection rules allow the following operator equation to hold
\begin{equation}\label{BEeqn}
\bar\CB_{1,r_3(0,0)}\ni\mu^{11}\cdot \CO_{r_3}=\kappa_1Q^{1\alpha}\CO_{r_1}\cdot Q^1_{\alpha}\CO_{r_2}+\kappa_2Q^{1\alpha}Q^1_{\alpha}\CO_{r_1}\cdot\CO_{r_2}+\kappa_3\CO_{r_1}\cdot Q^{1\alpha}Q^1_{\alpha}\CO_{r_2}\in\bar\CB_{1,r_1+r_2-1(0,0)}~,
\end{equation}
when $r_3=r_1+r_2-1$ (we have chosen a normalization in which $r(Q^i_{\alpha})=-1/2$).\footnote{Note that such a situation can never occur in a unitary rank one theory, i.e. a theory in which there is a single $\bar\CE_{r}$ generator of $\CA_{\CC}$ or of the Coulomb branch chiral ring, $\CR_{\CC}$. Indeed, in this case $r_i=n_ir$ with $n_i\in\mathbb{Z}$, and we have
\begin{equation}
r_3-r_1-r_2=r(n_3-n_1-n_2)\ne-1~.
\end{equation}
In deriving the above constraint, we have used the fact that the $n_i\in\mathbb{Z}$ and that $r>1$ by unitarity (in the free case we have $r=1$, but the RHS of \eqref{BEeqn} cannot be a primary).} Here the $\kappa_i\in\mathbb{C}$ are constants chosen to make the righthand side a superconformal primary, $\mu^{11}$ is the highest $\mathfrak{su}(2)_R$-weight primary of a $\hat\CB_1^{\rm neutral}$ multiplet, $\CO_{r_i}$ are $\bar\CE_{r_i}$ primaries, and \lq\lq$\cdot$" should be understood as a product in a $1/4$-BPS chiral ring.\footnote{While we are not aware of any fully quantum superconformal theories in which \eqref{BEeqn} holds non-trivially (i.e., where the operators on both sides of the equation are non-vanishing), the following example comes close. Consider the theory of a $U(1)$ vector multiplet coupled to a charged hypermultiplet via the superpotential term $W=g\phi q\tilde q$ (such couplings often occur in the IR effective theory on the Coulomb branch). Take $\CO_{r_1}=\CO_{r_2}=\phi$ (technically speaking, at $g=0$, $\phi$ is the primary of a $\bar\CD_{0(0,0)}$ multiplet, but we can think of such a superfield as an $\bar\CE_1$ representation). This theory is classically superconformal (but $g$ is irrelevant in the quantum theory). We then have \eqref{BEeqn} with $\kappa_1=0$, and the operator in \eqref{BEeqn} is a supersymmetric primary (in the sense that it cannot be written as the action of a Poincar\'e supercharge on a well-defined local operator).\label{irrCoupling}} Taking the OPE of both sides of \eqref{BEeqn} with $\CE_{-r_3}$ yields \eqref{BEEE}. Taking further products of the operators appearing in \eqref{BEeqn} can then, in principle (i.e., without null relations due to some dynamical constraint), give \eqref{BRint} for any integral $R$.

This discussion has an interesting echo in the phenomenon observed in \cite{Argyres:1995xn}. There the authors noted that in the RG flow from $\CN=2$ $SU(2)$ SQCD with $N_f=1$ to the MAD SCFT, the UV $U(1)$ mass parameter dual to $q\tilde q\in\hat\CB_1$ maps to the IR prepotential deformation dual to the $\bar\CE_{6/5}$ multiplet
\begin{equation}
m_{UV}\to\lambda_{IR}\sim m_{\rm UV}\cdot\Lambda^{-{1\over5}}~.
\end{equation}
In some sense, \eqref{BEEE} is a superconformal version of this kind of mixing at the level of operators instead of relevant couplings.

\subsec{Coulombic theories}\label{Coulombic}
As we gave emphasized repeatedly, the Coulomb branch algebra, $\CA_{\CC}$, is likely to be non-trivial in the case of any interacting 4d $\CN=2$ SCFT. Moreover, given that we have shown $\CA_{\CC}\le\CA_{\CO,+}$, it is interesting to ask what happens in 4d $\CN=2$ SCFTs that are \lq\lq extremal" with respect to this inclusion
\begin{equation}
\CA_{\CC}=\CA_{\CO,+}=\CA_{\CO}~,\ \ \ \CA_{\CO,-}=\emptyset~.
\end{equation}
Such theories have their local operator algebra, $\CA_{\CO}$, generated by the Coulomb branch operators. We call such theories \lq\lq Coulombic:"

\bigskip
\noindent
{\bf Definition 4:} A Coulombic 4d $\CN=2$ SCFT is an SCFT whose local operator algebra is generated by $\bar\CE$ and $\CE$ multiplets. For the purposes of this definition, we take $\bar\CD_{0(0,0)}$ to be a multiplet of type $\bar\CE_1$.

\medskip
 Two examples of such theories are as follows (in descending order of rigor):
 \begin{itemize}
 \item The free $\CN=2$ vector multiplet. Its local operator algebra is obviously generated by the vector multiplet itself, $\bar\CE_1\oplus\CE_{-1}$.
 \item The MAD theory (a.k.a. the $(A_1, A_2)$ SCFT) is conjectured to be an example of such an SCFT with generators $\bar\CE_{6/5}\oplus\CE_{-6/5}$ \cite{Bhargava:2022cuf,Bhargava:2022yik}. We discussed some evidence for this conjecture arising from the $\bar\CC$ spectrum of this theory in the $\bar\CC$ bullet in the list below \eqref{4dN2irreps}. We will explain further evidence in favor of this conjecture in the next section.
\end{itemize}

\medskip
Given this definition and the discussion in Sec. \ref{ACgen}, we arrive at the following theorem:

\bigskip
\noindent
{\bf Theorem 5:} In any Coulombic theory, the associated 2d VOA, in the sense of \cite{Beem:2013sza}, only has currents of integer 2d spin (i.e., $h\in\mathbb{Z}$).

\medskip
\noindent
As part of a broader suggested explanation of 4d/2d phenomenology in the next section, we will see that this theorem is satisfied for the free vector (see also the comments below corollary 3) and the MAD SCFT.

\newsec{Toward an explanation of VOA spins from the Coulomb branch algebra}\label{VOAspins}
In this section we further explore the link between Coulomb branch algebras and the 2d VOAs associated with local unitary 4d $\CN=2$ SCFTs. The spins (i.e., holomorphic scaling dimensions) and statistics of VOA currents are the most basic observables associated with the 4d/2d map in \cite{Beem:2013sza} (the spin-statistics theorem is violated due to the lack of unitarity in the 2d theory). In general, $\CA_{\CC}$ contains both fermionic and bosonic Schur operators, and so the Coulomb branch algebra gives rise to both bosonic and fermionic 2d VOA states (4d statistics translates directly into 2d statistics via the procedure in \cite{Beem:2013sza}).\footnote{The reason for this fact is that the Grassman parameters entering \eqref{3pt} are fermions.} On the other hand, as we have seen in Theorem 5, any Schur operator that lives in $\CA_{\CC}$ corresponds to a 2d VOA current of integer 2d spin. 

At the level of 2d spins, one of the most striking aspects of the phenomenology of 2d VOAs associated with 4d $\CN=2$ SCFTs is the following:

\bigskip
\noindent
{\bf Observation 6:} To the best of our knowledge, all known 2d VOAs, in the sense of \cite{Beem:2013sza}, corresponding to local unitary 4d $\CN=2$ theories lacking a Higgs branch\footnote{By Higgs branch, we mean any $\CN=2$ moduli space of vacua on which $\mathfrak{su}(2)_R$ is spontaneously broken (we do not require the low-energy theory on the moduli space to consist solely of hypermultiplets). \label{Higgs}} only have currents of integer 2d spin.

\medskip
\noindent
For example, consider the following theories, all of which lack Higgs branches:
\begin{itemize}
\item The free $\CN=2$ super Maxwell theory has an associated VOA consisting of the small algebra of a $(b,c)$ ghost system of weights $(1,0)$. Alternatively, we can think of the corresponding 2d theory as the $A(2)$ VOA \cite{Feigin:2007sp} (see the final bullet below for a 4d/2d map involving the closely related $A(6)$ VOA). This VOA has only integer 2d spin currents.
\item The MAD theory (a.k.a. the $(A_1, A_2)$ SCFT) has the Lee-Yang VOA as its associated 2d theory. Clearly, all states in this VOA have integer 2d spins (they are normal-ordered products of the EM tensor and its derivatives).
\item The $(A_1, A_{2n})$ SCFTs have $M(2,2n+3)$ Virasoro minimal model VOAs (the previous bullet corresponds to $n=1$) \cite{Cordova:2015nma}. Again, since the EM tensor generates this VOA, all currents have integer 2d spin.
\item The $(A_{N-1}, A_{M-1})$ SCFTs with $N$ and $M$ co-prime (the previous bullet corresponds to $N=2$ and $M=2n+1$) have as their VOAs the $W_N$-algebra minimal models of type $(N,N+M)$ \cite{Cordova:2015nma}. All the generating currents (and therefore all states in the VOA) have integer 2d spin.
\item The diagonal $SU(2)$ gauging of three $(A_1, A_3)$ theories described in \cite{Buican:2016arp,Buican:2020moo,Kang:2021lic,Jiang:2024baj}\footnote{Recall that each $(A_1, A_3)$ theory has an $\mathfrak{su}(2)$ flavor symmetry.} (in \cite{Buican:2020moo}, this theory is referred to as the \lq\lq $(3,2)$ SCFT"). We work {\it at non-zero $SU(2)$ gauge coupling}, $g\ne0$ (and away from any other cusps). In this case, the associated 2d theory is the $A(6)$ VOA \cite{Buican:2016arp}. The generators of this VOA are the EM tensor and two fermionic currents with $h=4$. Therefore, the corresponding VOA has only integer 2d spins. The reason we insist on working away from $g=0$ and other cusps is that the short multiplet spectrum and associated VOA change at these special points (see also \cite{Perlmutter:2020buo,Baume:2023msm}). For example, at $g=0$, the theory has non-trivial $\hat\CB_R$ multiplets arising from the $(A_1, A_3)$ factors (each corresponding VOA is a copy of $\widehat{\mathfrak{su}(2)}_{-4/3}$ \cite{Buican:2015ina,Buican:2015hsa}) and giving rise to a Higgs branch. As we turn on the gauge coupling, these  multiplets recombine to form long multiplets (and the Higgs branch is lifted; see \cite{Argyres:2012fu} for a similar phenomenon on the moduli space of certain $\CN=2$ SYM theories). Note that, even at $g=0$, the associated VOA only has integer scaling dimensions. This statement follows from the fact that the $(A_1, A_3)$ theories have integral spin VOAs and the free vectors that we add to the theory to gauge the $SU(2)$ also have Schur operators of integral 2d spin.\footnote{Therefore, we see that Observation 6 does not lead to a necessary condition for an associated VOA to have integral spin.} 
\end{itemize}

We can connect our observation to the discussion in the previous sections by noting that, for SCFTs that lack a Higgs branch (in the sense of footnote \ref{Higgs}), it is reasonable to conjecture that the Coulomb branch multiplets generate the local operator algebra, $\CA_{\CC}=\CA_{\CO}$. The heuristic motivation is that if the only moduli space of vacua is the Coulomb branch, then we expect to be able to reconstruct the SCFT at the origin by studying the Coulomb branch effective theory and certain simple deformations thereof. We can think of the statement that $\CA_{\CC}=\CA_{\CO}$ as a version of this intuition at the origin of the moduli space.

\bigskip
\noindent
{\bf Conjecture 7:} All of the theories in the above list are Coulombic.\footnote{As discussed in the previous section, the free vector theory is Coulombic. Moreover, we described non-trivial evidence from the $\bar\CC$ spectrum suggesting that the MAD theory is also Coulombic (see \cite{Bhargava:2022cuf,Bhargava:2022yik} for a more detailed discussion of this evidence and the associated conjecture). In light of theorem 5, the fact that the associated VOA has integer 2d spin is further evidence for the conjecture that the MAD theory is Coulombic. In fact, using the results of \cite{Bhargava:2024yqv}, we can strengthen this argument. In this latter paper, we showed that the Lee-Yang vacuum character can be understood as arising from the fact that the following MAD normal-ordered product vanishes: $J^{11}_{+\dot+}J^{11}_{+\dot+}=0$ (where $J^{11}_{+\dot+}$ is the highest $\mathfrak{su}(2)_R\oplus\mathfrak{so}(4)$-weight component of the $\mathfrak{su}(2)_R$ current). Indeed, this result strongly suggests that
\begin{equation}
\CA_{\rm Schur}^{\rm MAD}<\CA^{\rm MAD}_{\CC}~.
\end{equation}
} More generally, local unitary 4d $\CN=2$ SCFTs lacking a Higgs branch (in the sense of footnote \ref{Higgs}) are Coulombic (for theories that have a marginal coupling, we assume we work at a generic point on the conformal manifold away from cusps or other special points\footnote{Note that at the $g=0$ point of the $(3,2)$ conformal manifold, the theory cannot be Coulombic: the corresponding $\hat\CB_R$ multiplets cannot appear in the OPEs of the vector multiplets we add to gauge the $SU(2)$. Moreover, operators like the $\hat\CB_3$ multiplet corresponding to the $g=0$ Schur operator $\CO_{abc}:=\epsilon_{IJK}\mu_1^I\mu_2^J\mu_3^K$ cannot appear in any of the $\CA_{\CC}$ algebras of the individual $(A_1, A_3)$ SCFTs (here the subscript of $\mu^I_i$ denotes the particular $(A_1, A_3)$ factor out of the three present, and $I,J,K$ are $\mathfrak{su}(2)$ adjoint gauge indices). Indeed, the reason is that each such operator is charged under the individual $\mathfrak{su}(2)$ symmetries of a given $(A_1, A_3)$ factor, but we have seen that Coulomb branch operators are uncharged under flavor Lie algebras.}).

As evidence in favor of our conjecture, note that, if it is true, then Theorem 5 gives a particularly simple explanation of Observation 6. In particular, the $\CI$ symmetry gives us a powerful organizing principle that explains a very basic piece of phenomenology. An appealing aspect of this conjecture is that it also implies (assuming the discussion in \cite{Beem:2017ooy} that relates the 4d Higgs branch with the 2d associated variety of the VOA arising via the map in \cite{Beem:2013sza}) the following closely related conjecture which can be phrased mostly in 2d language:

\bigskip
\noindent
{\bf Conjecture 7$'$:} Any 2d VOA that has a trivial associated variety and is related to a unitary local 4d $\CN=2$ SCFT via \cite{Beem:2013sza} necessarily only has VOA states of integer 2d spin.

\medskip
\noindent
Note that, in the above conjecture, it is crucial that we insist on a link to 4d. Otherwise, the 2d $\CN=1$ Super Virasoro minimal models are counterexamples.\footnote{We thank S.~Wood for this observation.} The fact that, upon going to 4d via \cite{Beem:2013sza}, these theories violate CPT is perhaps a hint of a more general strategy to prove this conjecture.

\newsec{Theories with $\CN>2$ SUSY}\label{Ng2}
Let us now focus on local unitary SCFTs with $\CN>2$ SUSY and explain how the above discussion can be generalized. First, recall from Corollary 3 in Sec. \ref{SchurSec}, that any such theory has faithful $\CI$ since the extra supercurrents (from the manifest $\CN=2$ perspective) sit in multiplets satisfying
\begin{equation}\label{FaithfulNg2}
\CI(\bar D_{1/2(0,0)})=-1~.
\end{equation}

We will specialize to $\CN=4$ theories for the time being and comment further on $\CN=3$ SCFTs at the end of this section. From an $\CN=2$ perspective, the extra $\CN=4$ supercurrent multiplets transform as fundamentals of an $\mathfrak{su}(2)_{F}$ flavor symmetry that commutes with the manifest $\CN=2$ superconformal algebra. Let us denote the center of the corresponding simply connected group, $SU(2)_F$, as $\CF:=Z(SU(2)_F)\cong\mathbb{Z}_2$. Then, we have
\begin{equation}
\CF(\bar D^a_{1/2(0,0)})=-1\ \Rightarrow\ \hat\CI(\bar D^a_{1/2(0,0)})=1~,\ \hat\CI:={\rm Diag}(\CI\times\CF)~,
\end{equation}
and so we can imagine $\CN=4$ SCFTs that are neutral under $\hat\CI$ even though they are not neutral under $\CI$.\footnote{We can alternatively define $\hat\CI:={\rm Diag}(Z(SU(4)_R)|_{\mathbb{Z}_2}\times Z(Spin(4)))$, where $SU(4)_R\times Spin(4)$ is the simply connected group associated with $\mathfrak{su}(4)_R\oplus\mathfrak{so}(4)<\mathfrak{psu}(2,2|4)$. Here we use the $\mathbb{Z}_2<\mathbb{Z}_4\cong Z(SU(4)_R)$ subgroup of the $R$-symmetry group center in defining $\hat\CI$.}

In fact, all known 4d $\CN=4$ SCFTs are of Super Yang-Mills (SYM) type with some gauge group, $G$, and therefore have unfaithful $\hat\CI$ (this situation is analogous to that described in previous sections for the free $\CN=2$ vector and $\CI$). Indeed, it is easy to see that 

\begin{eqnarray}
\CI(\Phi^A)&=&1~,\ \ \CI(q^{A}_f)=\CI(\tilde q^A_f)=-1~, \cr \CF(\Phi^A)&=&1~,\ \ \ \CF(q_f^A)=\CF(\tilde q_f^A)=-1~,\cr \hat \CI(\Phi^A)&=&\hat\CI(q_f^A)=\hat\CI(\tilde q_f^A)=+1~,
\end{eqnarray}
where $A$ is a $G$ adjoint index, and $f$ is a fundamental index of $\mathfrak{su}(2)_F$. Here $\Phi^a$ is the scalar of the $\CN=2$ vector submultiplet of the $\CN=4$ gauge multiplet, and $q_f$ and $\tilde q_f$ are the scalars of the $\CN=2$ adjoint hypermultiplets that form the remaining $\mathfrak{su}(4)_R$ states of the $\CN=4$ vector multiplet primary. By construction, the fermions and the field-strength tensor sitting inside the $\CN=4$ vector multiplet also have $\hat\CI=1$. As a result, the $\CN=4$ vector multiplet is neutral under $\hat\CI$, and so we see that all local operators in any $\CN=4$ SYM theory are necessarily neutral
\begin{equation}
\hat\CI(\CO)=1~,\ \ \ \forall\ \CO\in\CH_L^{\CN=4\ SYM}~.
\end{equation}
At the level of the associated 2d VOA, we have that $h-j_F\in\mathbb{Z}$, where $j_F$ (satisfying $|j_F|\ge1/2$) is the weight under the $\mathfrak{sl}(2)_R$ of the $\mathfrak{psl}(2|2)$ part of the small 2d $\CN=4$ superconformal algebra (in \cite{Bonetti:2018fqz}, this 2d $R$ symmetry is referred to as $\mathfrak{sl}(2)_y$). Summarizing this logic, we have:

\bigskip
\noindent
{\bf Fact 8:} All known $\CN=4$ SCFTs (i.e., SYM theories) have $\hat\CI=1$ on the space of local operators, and their associated 2d $\CN=4$ VOAs, in the sense of \cite{Beem:2013sza}, have $h-j_F\in\mathbb{Z}$.

\medskip
Is this statement true more generally? In principle, there are representations of the $\CN=4$ SCA that cannot be built from SYM fields for any $G$ and hence can only exist in hypothetical \lq\lq exotic" $\CN=4$ SCFTs. In analogy with \eqref{Apm}, we can classify $\CN=4$ SCA irreps (in the nomenclature of \cite{Cordova:2016emh}) with respect to the $\hat\CI$ charge. For simplicity, we only discuss multiplets with $\hat\CI=-1$ (we also drop multiplets related to those appearing below by CPT)
\begin{eqnarray}\label{ApmN4}
\CA_{\CO,-}^{\CN=4}&\ni& L\bar B_1[2j;0]^{(R_1, R_2,R_3)}|_{R_1+R_3+2j\in\mathbb{Z}_{\rm odd}}~,\ L\bar A_2[2j;0]^{(R_1, R_2,R_3)}|_{R_1+R_3+2j\in\mathbb{Z}_{\rm odd}}~,\cr&&\ L\bar A_1[2j;2\bar j\ge1]^{(R_1, R_2,R_3)}|_{R_1+R_3+2j+2\bar j\in\mathbb{Z}_{\rm odd}}~,\ L\bar L[j;\bar j]^{(R_1,R_2,R_3)}|_{R_1+R_3+2j+2\bar j\in\mathbb{Z}_{\rm odd}}~,\ \ \ \ \ \ 
\end{eqnarray}
where $(R_1,R_2,R_3)$ are Dynkin labels for $\mathfrak{su}(4)_R$,
 and $\hat\CI$ is generated by
\begin{equation}\label{IhatGen}
\rho_{\hat\CI}:=e^{\pi i(R_1+R_3+2j+2\bar j)}~.
\end{equation}

None of the multiplets in \eqref{ApmN4} can appear in an $\CN=4$ SYM theory.\footnote{Depending on what one means by an $\CN=4$ SYM theory, there is a possible loophole in this argument. The reason is that, a priori, we are not yet able to rule out the existence of $\CN=4$ SYM theories with 2-form symmetries that give rise to a faithful $\hat\CI$ upon gauging (or, in the condensed matter language, \lq\lq condensing" the topological lines that generate the 2-form symmetry). Note that $\hat\CI$ would then be a corresponding \lq\lq quantum" 0-form symmetry, and local operators with $\hat\CI=-1$ would appear. In this case, we would have a sector of $\CN=4$ SYM operators that cannot be written in terms of the $\CN=4$ vector multiplet fields. Such a theory would therefore share the $\CA_{\CO,+}^{\CN=4}$ operator algebra with \lq\lq standard" $\CN=4$ SYM theories but would also contain an \lq\lq exotic" sector. This sector would be admittedly peculiar: any non-vanishing normal-ordered product $:\CO_1\CO_2:$ with $\CO_{1,2}\in\CA_{\CO,-}^{\CN=4}$ would be in $\CA_{\CO,+}^{\CN=4}$ and could be written in terms of SYM vector multiplets. Then, going to a cusp would give us a way to write $:\CO_1\CO_2:$ in terms of free fields. Such a scenario would also give us two SCFTs with different local operator content that have the same 2d VOA via \cite{Beem:2013sza} (to our knowledge, there are no such examples in the literature).

Note that the discrete gauging described in \cite{Argyres:2018wxu} gives rise to a $\mathbb{Z}_2$ 2-form symmetry. However, condensing the corresponding lines gives a 0-form symmetry $\mathbb{Z}_2:={\rm Diag}(Z(SU(4)_R)\times Z(SL(2,\mathbb{Z}))\ne\hat\CI$.\label{2form}} Moreover, they only satisfy shortening conditions with respect to the $\bar Q^{\bar I}_{\dot\alpha}$ supercharges (here $\bar I$ is an anti-fundamental index of $\mathfrak{su}(4)_R$) and not the $Q^I_{\alpha}$ supercharges (with the opposite statement holding for the CPT-conjugate multiplets). Indeed, from an inspection of the irreps in \cite{Cordova:2016emh}, it turns out that any multiplet satisfying shortening conditions with respect to both sets of supercharges is necessarily in $\CA_{\CO,+}^{\CN=4}$.

Using the above analysis, one can check the following statement:

\bigskip
\noindent
{\bf Theorem 9:} None of the multiplets in the $\CA_{\CO,-}^{\CN=4}$ sector correspond to $\CN=4$ irreps containing $\CN=2$ Coulomb branch or Schur operators. In particular, the latter statement implies that $\CA_{\CO,-}^{\CN=4}$ does not contribute states to the associated 2d VOA.

\medskip
\noindent
{\bf Proof:} The statement is true for Coulomb branch multiplets since we have already seen they are trivial under $\CI$ and, moreover, they are neutral under $\mathfrak{su}(2)_F$ by the logic around \eqref{ChirOPE}.

Next, consider the Schur multiplets. We have argued that, by inspection of the quantum numbers in \cite{Dolan:2002zh} (see also \cite{Cordova:2016emh}), the only possible multiplets (up to the action of CPT) in $\CA_{\CO,-}^{\CN=4}$ obey $\bar Q$ shortening conditions but are \lq\lq long" with respect to the $Q$ supercharges. Since $\CN=2$ Schur multiplets satisfy shortening conditions with respect to both chiral and anti chiral $\CN=2<\CN=4$ supercharges, we see they cannot live in such multiplets. Therefore, the multiplets in $\CA_{\CO,-}^{\CN=4}$ do not contribute to the associated 2d VOA. $\square$

\medskip
\noindent
This theorem also leads to a simple but potentially constraining corollary from the 2d VOA perspective:

\bigskip
\noindent
{\bf Corollary 10:} If any 2d VOA with small $\CN=4$ supersymmetry has states with $h-j_F\not\in\mathbb{Z}$, then the VOA does not have a 4d parent (in the sense of \cite{Beem:2013sza}).\footnote{Note that 2d VOA states with $h-j_F\not\in\mathbb{Z}$ are necessarily in long multiplets of $\mathfrak{psl}(2|2)$. In \cite{Bonetti:2018fqz}, the authors find long multiplets in 2d VOAs corresponding to certain $\CN=4$ SYM theories. However, these long multiplets all have $h-j_F\in\mathbb{Z}$.}

\medskip
Our corollary therefore gives constraints, for arbitrary two-dimensional central charge $c_{2d}$, that VOAs related to 4d $\CN=4$ SCFTs must satisfy. It would be interesting to find further constraints that such VOAs obey. 

Our discussion also motivates the following conjecture:

\bigskip
\noindent
{\bf Conjecture 11:} All local unitary $\CN=4$ SCFTs have an unfaithful action of $\hat\CI$ on the space of local operators. In other words, all such theories have $\CA_{\CO,-}^{\CN=4}=\emptyset$.

\medskip
\noindent
Indeed, otherwise the SCFT in question would be very unusual: multiplets in $\CA_{\CO,-}^{\CN=4}$ would not contribute to any correlation functions involving $\CN=4$ multiplets containing $\CN=2$ Coulomb branch and Schur sector multiplets. However, see footnote \ref{2form} for potential sources of counterexamples (in some sense, it would be more interesting for this conjecture to be wrong).

Finally, let us comment on $\CN=3$ SCFTs. By the discussion around \eqref{FaithfulNg2}, $\CI$ also acts faithfully on the space of local operators of these theories. We can try to define an analog of the unfaithful $\hat\CI$ for $\CN=3$ SCFTs (in an attempt to identify \lq\lq exotic" $\CN=3$ SCFTs)  by mixing $\CI$ with $\CF_3\cong\mathbb{Z}_2<U(1)_{F_3}$, where $\mathfrak{u}(1)_{F_3}<\mathfrak{su}(3)_R\oplus \mathfrak{u}(1)_{r_3}$ is a subalgebra of the $\CN=3$ $R$ symmetry that is a flavor symmetry from the $\CN=2$ perspective. To have the properties we want, we should require that
\begin{equation}
\CF_3(\bar D_{1/2(0,0)})=-1\ \Rightarrow\ \CI_3(\bar D^a_{1/2(0,0)})=1~,\ \CI_3:={\rm Diag}(\CI\times\CF_3)~.
\end{equation}
Without loss of generality, we assume that the extra chiral supercharge, $Q'_{\alpha}$, has charge $-1$ under $U(1)_{F_3}$ (so that $\CF_3$ is generated by $\exp(\pi i)\in U(1)_{F_3}$). We will also assume that all other local operators have integer $F_3$ charge.\footnote{We do not know if such a condition is generally satisfied by $\CN=3$ SCFTs.} Otherwise, $\CI_3$ will give rise to (model-dependent) complex phases when it acts on some of the local operators (equivalently, the $U(1)_{r_3}$ symmetry will participate in an extension specified by the spectrum of charges\footnote{See \cite{Dedushenko:2023cvd} for a discussion of a similar phenomenon in purely $\CN=2$ theories.}). Using our above assumptions, we obtain the generator of $\CI_3$
\begin{equation}
\rho_{\CI_3}:=e^{\pi i\left({4\over3}R_1+{2\over3}R_2+{1\over3}r_3+2j+2\bar j\right)}~.
\end{equation}

We can now use the list of $\CN=3$ irreps in \cite{Cordova:2016emh} to conclude that (up to CPT) the multiplets with $\CI_3=-1$ are (see also \cite{Lemos:2016xke}; note that our $j$ and $\bar j$ correspond to spins and are therefore half the corresponding Dynkin labels used in \cite{Cordova:2016emh})
\begin{eqnarray}\label{AN3def}
\CA_{\CO,-}^{\CN=3}&\ni&L\bar B_1[2j;0]^{(R_1, R_2;r_3>r_*+6)}|_{{4\over3}R_1+{2\over3}R_2+{1\over3}r_3+2j\in\mathbb{Z}_{\rm odd}}~,\ L\bar A_2[2j;0]^{(R_1, R_2;r_3>r_*)}|_{{4\over3}R_1+{2\over3}R_2+{1\over3}r_3+2j\in\mathbb{Z}_{\rm odd}}~,\cr&&\ L\bar A_1[2j;2\bar j\ge1]^{(R_1, R_2;r_3>r_*)}|_{{4\over3}R_1+{2\over3}R_2+{1\over3}r_3+2j+2\bar j\in\mathbb{Z}_{\rm odd}}~,\ \cr&&\ L\bar L[j;\bar j]^{(R_1,R_2;r_3)}|_{{4\over3}R_1+{2\over3}R_2+{1\over3}r_3+2j+2\bar j\in\mathbb{Z}_{\rm odd}}~,\cr r_*&:=&6(j-\bar j)+2(R_1-R_2)~.
\end{eqnarray}
Note that, as in the $\CN=4$ case, there are no contributions to $\CA_{\CO,-}^{\CN=3}$ from multiplets satisfying shortening conditions with respect to both chiral and anti-chiral supercharges. Indeed, up to CPT, the multiplets in $\CA_{\CO,-}^{\CN=3}$ only satisfy shortening conditions with respect to the chiral $\CN=3$ supercharges. Therefore, we have a version of Theorem 9 that applies to $\CN=3$ SCFTs (it can be proven in the same way as its $\CN=4$ analog):

\bigskip
\noindent
{\bf Theorem 12:} Subject to the assumption that the extra $Q'_{\alpha}$ and $\bar Q'_{\dot\alpha}$ $\CN=3$ supercharges have minimal non-zero magnitude $U(1)_{F_3}$ charge, $\CA_{\CO,-}^{\CN=3}$ exists and takes the form in \eqref{AN3def}. Moreover, $\CA_{\CO,-}^{\CN=3}$ does not contain $\CN=3$ irreps housing $\CN=2$ Coulomb branch or Schur multiplets. In particular, the latter statement implies that $\CA_{\CO,-}^{\CN=3}$ does not contribute states to the associated 2d VOA.

\newsec{Discussion}
In this paper, we have empasized the role the $\CI$ symmetry plays in providing universal selection rules for the local operator algebra and $n$-point functions of 4d $\CN=2$ SCFTs. For example, we have argued that it gives important constraints on the Coulomb branch operator algebra, $\CA_{\CC}$, (and related contributions to $n$-point functions of $\bar\CE$ and $\CE$ multiplets) and helps explain certain phenomena in the 4d/2d correspondence of \cite{Beem:2013sza}.

We suspect that better understanding the $\CI$ symmetry (e.g., its gaugings and the 't Hooft anomalies it participates in) will shed additional light on the global structure of 4d $\CN=2$ SCFTs and theories connected to them via continuous deformations. We also expect that the $\CI$ symmetry and various related symmetries (like the $\hat\CI$ symmetry we described in the $\CN=4$ case) will provide constraints for bootstrapping and eventually solving SCFTs with extended supersymmetry in 4d. We hope that proving (or disproving) some of our conjectures will help.

We conclude with a list of open questions and further ideas:
\begin{itemize}
\item Another natural operator (sub)-algebra to study is the combined Coulomb / Schur operator algebra
\begin{equation}\label{CSalgebra}
\CA_{\CC S}:=\CA_{\CC}\times \CA_{\rm Schur}~.
\end{equation}
What we mean by the above expression is the operator algebra generated by the Coulomb branch multiplets and the Schur multiplets.\footnote{Note that in the case of Coulombic theories, $\CA_{\rm Schur}<\CA_{\CC}$ and so $\CA_{\CC S}=\CA_{\CC}$. More generally, this statement need not hold.} This combined operator algebra characterizes many important properties of 4d $\CN=2$ SCFTs.\footnote{Note that Coulomb branch operators and Schur operators, while describing different physics, \lq\lq know" a lot about each other (for some initial results in this direction, see \cite{Buican:2015hsa,Fredrickson:2017yka}; more recently see also \cite{Shan:2023xtw}). It would be interesting to understand this relationship in full generality.} In particular, it would be interesting to understand when this combined algebra saturates the local operator algebra of the SCFT, $\CA_{\CC S}=\CA_{\CO}$ (i.e., when the Coulomb branch and Schur sector operators generate the local operator algebra).\footnote{In Coulombic theories, $\CA_{\rm Schur}<\CA_{\CC}=\CA_{\CO}$ and so $\CA_{\CC S}=\CA_{\CO}$.} Note that this algebra is sensitive to the continuous flavor symmetries of 4d $\CN=2$ SCFTs (unlike $\CA_{\CC}$), although, a priori, it is logically possible that it may sometimes fail to capture the global form of the continuous flavor symmetry of $\CA_{\CO}$.

As a final note regarding \eqref{CSalgebra}, we observe that it is natural to relate $\CA_{\CC S}$ to the structure of moduli spaces, and it would be intriguing to understand if this algebra can be used to give a more intrinsic SCFT definition of the moduli space. Note that the Schur sector naturally captures properties of the Higgs branch, and so we expect all moduli spaces of 4d $\CN=2$ SCFTs are described by operators in \eqref{CSalgebra}. 
\item Although we have not done so here, it is also natural to construct (necessarily more model-dependent) selection rules for the local operator algebra using $\mathfrak{u}(1)_R<\mathfrak{su}(2,2|2)$ as well. We expect such selection rules to be particularly powerful when we have Coulomb branch operators of fractional scaling dimension (as in Argyres-Douglas theories). In such cases, the $U(1)_R$ symmetry participates in a group extension that includes discrete flavor symmetries (e.g., see \cite{Dedushenko:2023cvd}). Such symmetries should give particularly strong constraints on $\CA_{\CC}$ (note, however, that our inclusion in \eqref{BEEE} is compatible with this scenario due to the condition on the $\mathfrak{u}(1)_R$ charges).
\item In the case of $\CN=4$ SYM theories, we saw that $\CI$ acted faithfully on the local operator algebra but that we could deform $\CI$ by mixing with $Z(SU(2)_F)$ to produce an unfaithful $\hat\CI$ symmetry. In the context of more general $\CN=2$ SCFTs, we can ask if an analogous deformation is always possible. For example, in the case of $\CN=2$ $SU(2n+1)$ SQCD we have a faithful $\CI$ action on baryons, but we can mix with $\mathbb{Z}_2^B<U(1)_B$ to produce a symmetry $\CI_B:={\rm Diag}(\CI\times\mathbb{Z}_2^B)$ that acts unfaithfully on the local operators. In general, such deformations are unlikely to exist (e.g., see the examples in \cite{Distler:2020tub}), but, in strongly coupled SCFTs, it is challenging (and model-dependent) to find the full spectrum of discrete symmetries.
\item As discussed briefly in the introduction, there is an interesting connection between unfaithful $\CI$ (and more general unfaithful mixings described in the previous bullet) and the ability to place the corresponding theories on non-spin manifolds \cite{Cordova:2018acb,Brennan:2022tyl,Aspman:2022sfj}. In particular, the authors of \cite{Cordova:2018acb,Brennan:2022tyl, Aspman:2022sfj} argued that $\CN=2$ $SU(2)$ SYM and SQCD can be placed on such more general manifolds. In the case of SQCD this statement holds even though $\CI$ is non-trivial when acting on non-genuine local operators (e.g., the squark superfields, which we can think of as sitting at the end of Wilson lines). It is tempting to argue that all the theories with unfaithful $\CI$ we are considering here can also be placed on more general manifolds,\footnote{For example, it is tempting to conjecture that all Coulombic theories can be placed on non-spin manifolds (this statement is certainly true for the free vector and also follows for the MAD theory from the analysis in \cite{Brennan:2022tyl, Aspman:2022sfj}).} but, at the level of abstraction we are discussing such SCFTs (e.g., we do not assume that they are connected to free theories in the UV), it is not yet clear to us how to answer this question precisely. Along the same lines, it would be interesting to understand the 't Hooft anomalies involving $\CI$ (and its mixings) and also what theories we can obtain by gauging $\CI$ (when it is possible to do so).
\end{itemize}
We hope to return to some of these questions soon.

\ack{I am grateful to A.~Antinucci, A.~Banerjee, C.~Bhargava, P.~Genolini, H.~Jiang, T.~Nishinaka, L.~Tizzano, and S.~Wood for comments and discussions. I am also grateful to C.~Bhargava, H.~Jiang, and T.~Nishinaka for collaboration on a closely related project \cite{2024paper}. My work is supported by the Royal Society under the grant \lq\lq Relations, Transformations, and Emergence in Quantum Field Theory" and by STFC under the grant \lq\lq Amplitudes, Strings and Duality." No new data were generated or analyzed in this study.}

\newpage
\bibliography{chetdocbib}

\end{document}